\documentclass[letterpaper]{article} 
\usepackage{aaai2026}  
\usepackage{times}  
\usepackage{helvet}  
\usepackage{courier}  
\usepackage[hyphens]{url}  
\usepackage{graphicx} 
\usepackage{natbib}  
\usepackage{caption} 
\usepackage{algorithm}
\usepackage{algorithmic}

\usepackage[table]{xcolor}
\usepackage{mdframed}
\usepackage{longtable}
\usepackage{multirow}
\usepackage{colortbl}  
\usepackage{booktabs} 
\definecolor{darkgreen}{RGB}{102, 170, 104}
\definecolor{lightgreen}{RGB}{193, 225, 193}
\definecolor{lightred}{RGB}{240, 160, 160}
\definecolor{darkred}{RGB}{220, 100, 100}
\usepackage{newfloat}
\usepackage{listings}
\DeclareCaptionStyle{ruled}{labelfont=normalfont,labelsep=colon,strut=off} 
\floatstyle{ruled}
\newfloat{listing}{tb}{lst}{}
\floatname{listing}{Listing}
\title{``\textit{I want to be pushed, I want to grow}'': Enabling social workers to \\design evaluations of LLM augmentation in their work}
\author {
    Anna Kawakami\textsuperscript{\rm 1},
    Chloe Qianhui Zhao\textsuperscript{\rm 1},
    Renee Shelby\textsuperscript{\rm 2},\\
    Fernando Diaz\textsuperscript{\rm 1},
    Haiyi Zhu\textsuperscript{\rm 1},
    Kenneth Holstein\textsuperscript{\rm 3}
}
\affiliations {
    \textsuperscript{\rm 1}Carnegie Mellon University,
    \textsuperscript{\rm 2}Google Research, 
    \textsuperscript{\rm 3}EPFL\\
    akawakam@andrew.cmu.edu, qianhuiz@andrew.cmu.edu, 
    reneeshelby@google.com,\\diazf@cmu.edu, haiyiz@andrew.cmu.edu, ken.holstein@epfl.ch
}

\usepackage{bibentry}

\begin{document}

\maketitle

\begin{abstract}
Workers are increasingly asked to adopt AI systems to assist their work, yet are rarely given a voice in defining what meaningful AI augmentation should look like or how to evaluate for it. 
In this paper, we propose \textit{worker-driven AI measurement}---a bottom-up approach to AI evaluation where workers collaboratively shape decisions about which tasks AI should augment, what ``successful’’ augmentation looks like, and how it should be measured. We explore how to support this through a case study with 19 workers from a local school social work organization. Through a series of eight workshops, workers iteratively develop their own measurement goals for AI evaluation, systematize these goals, and then
design a benchmark to capture how effectively an LLM can ``challenge'' them to reflect on their own assumptions and biases in the context of their day-to-day work. Workers collaboratively design and refine an LLM-as-a-judge rubric based on their professional and lived expertise. In validations of the worker-created benchmark, we find that there is strong agreement between worker and LLM judge ratings and that the resulting benchmark can differentiate performance across six state-of-the-art LLMs. Based on our case study, we discuss opportunities for future work to support worker-driven AI measurement as a complementary approach to existing top-down AI evaluation approaches.   
\end{abstract}

\section{Introduction}
Frontline workers across occupations—from teaching and social work, to healthcare and finance—are increasingly asked to adopt AI systems to assist their work~\cite{brynjolfsson2025generative}. Yet, workers themselves rarely have a voice in defining what it would look like for AI to positively augment their day-to-day work, and how to design and evaluate for it~\cite{kawakami2026ai}. Today, decisions about which parts of workers’ work to augment, how to augment them, and how to measure what ``good performance’’ looks like are often left to technical experts or organizational leaders. Without worker involvement in these decisions, AI evaluations often reflect impoverished understandings of workers’ day-to-day work and the forms of expertise involved. For example, in occupations like social work and teaching, a lack of worker involvement has often led to AI deployments that conflict with worker values and practices~\cite{kawakami2026ai}. 

A growing body of work focuses on developing AI evaluations for specific occupations~\cite{budler2025brief}. For example, NLP and ML researchers have proposed new AI measurement instruments, such as benchmarks and metrics to measure the \textit{automatability} of specific work tasks in healthcare, software engineering, finance, and design (e.g.,~\cite{vishwakarma2025can,bedi2025medhelm,wang2025ai}). More recently, researchers have begun to recognize the need to involve workers themselves in shaping AI evaluations~\cite{bedi2025medhelm,arora2025healthbench,gebreegziabher2025metricmate}. Research in NLP and ML has explored methods to solicit feedback from workers to inform the design of AI evaluations. However, this work has typically constrained worker involvement to narrow sub-activities such as developing test cases or providing expert labels for benchmarks, without engaging workers in larger decisions such as what is worth measuring in the first place~\cite{kawakami2026measure}. By contrast, HCI research has solicited workers' perspectives 
on how new LLM-based systems should augment their professions in the future~\cite[e.g.,][]{brachman2025current,woodruff2024knowledge} ,
but has not 
directly engaged them in the design of concrete measurement instruments. 
\textbf{Taken together, it remains an open question how to meaningfully support workers in the end-to-end, collaborative design of AI measurement instruments}—from \textit{defining} what positive AI augmentation looks like from their perspective, to \textit{operationalizing} practical measurement instruments that can accurately capture these notions.

In this paper, we present a case study with a local school social work organization that is actively exploring how to integrate LLMs into their work. In partnership with workers at this organization, we explore an approach to \textit{worker-driven AI measurement design}—where workers collaboratively shape decisions around \textit{which tasks} AI should augment, \textit{what} ``successful’’ augmentation looks like for those tasks, and \textit{how} it should be measured. In particular, through an iterative series of eight workshops with 19 workers, we explore how to engage workers in (1) identifying realistic LLM use cases for which workers actually want LLM-based support, (2) developing a shared understanding of what ``good’’ or ``bad’’ LLM performance looks like for a given use case, and (3) operationalizing this understanding through the design of a benchmark (consisting of an LLM-as-a-judge system and a set of test cases). 

Overall, through these design workshops, workers collaboratively developed an LLM-as-a-judge system to evaluate model performance on a specific use case core to their work—developing useful reflection questions to guide future meetings with teachers, based on workers’ observations of teacher practices and classroom challenges. Workers first collaboratively designed 16 realistic test cases that they believed would be challenging for an LLM, based on their past experience using LLMs for this use case. In early workshops, workers identified a measurement goal that is largely absent in existing industry AI evaluations: 
measuring how effectively LLMs can function as \textbf{``people \textit{challengers}''} (a play on current LLM tendencies to be ``people pleasers.’’) In the context of their work, this meant challenging workers to \textit{`` reflect on their own possible assumptions and biases [and helping] them expand their perspectives, providing opportunities for learning and growth.’’} 

Workers iteratively operationalized this goal into an LLM-as-judge rubric  that captures their notions of what it looks like to meaningfully augment their work in their specific organizational and work context.  
Our validation shows that the resulting LLM-as-a-judge system has strong agreement with worker judgments, 
and that the benchmark can differentiate the performance of different state-of-the-art LLMs. 
Overall, we make the following contributions:
\begin{enumerate}
    \item We propose \textbf{\textit{worker-driven AI measurement}}—a bottom-up, collaborative design approach that complements existing evaluation practices for AI in work. 
    \item We explore this approach through an early \textbf{case study} with a local social work organization.
    \item We present \textbf{opportunities for future work} to support worker-driven AI measurement design across worker and organizational contexts.
\end{enumerate}

\section{Background} \label{background}
In this section, 
we overview past work that brings social science measurement concepts to AI evaluation (Section~\ref{background_AIeval}). 
We then discuss nascent efforts to engage workers in shaping AI evaluations (Section~\ref{background_workerp}). 

\subsection{AI Evaluation from a Social Science Measurement Perspective}\label{background_AIeval}
Following recent calls in responsible computing and machine learning, in this paper we understand AI evaluation as a \textit{social science measurement challenge}~\cite{wallach2025position,adcock2001measurement,jacobs2021measurement,johnson2026evaluating,kawakami2026measure}. AI evaluations necessarily rely on assumptions about both \textit{what} is important to measure and \textit{how} unobserved, latent constructs (e.g.,``fairness,’’ ``accuracy’’, ``safety’’) should be systematized and operationalized into practical measurement instruments, such as metrics and benchmarks. These assumptions often clash with reality in the downstream contexts where AI applications are used~\cite{jacobs2021measurement}. For example, developers and deployers make claims about the ``utility’’ of healthcare LLMs using benchmarks based on medical exams~\cite[e.g.,][]{wang2024mmlu,pal2022medmcqa,yao2024medqa} that primarily measure an LLM’s ability to \textit{recall} medical information~\cite{raji2025s}. However many of the tasks LLMs are used for in real-world clinical settings rely on other capabilities, and do not necessarily involve the recollection of medical information~\cite{raji2025s}. 
In fact, researchers have found that the majority of LLM benchmarks in the literature have poor construct validity~\cite[e.g.,][]{bean2025measuring}. Others note that existing LLM benchmarks and metrics tend to jump from identifying broad measurement goals (e.g., measuring ``harmfulness’’) to operationalization, without articulating or critically interrogating how the underlying concept should be defined in the first place~\cite{wallach2025position}. 

While prior work has begun to offer shared language and frameworks to ground AI evaluation in practices from social science measurement theory, comparatively little work has focused on how to meaningfully involve non-technical stakeholders in the design of measurement instruments for evaluating LLMs. 
Recently, researchers have drawn from ideas from measurement theory to explore how to engage non-AI-expert stakeholders in shaping LLM evaluations~\cite{johnson2026evaluating, kawakami2026measure,smith2024recommend}---for example, by engaging impacted communities in systematizing the concept of ``cultural appropriateness’’ for evaluating text-to-image models~\cite{johnson2026evaluating}. 

To date, past work has explored how to support stakeholder participation in specific, isolated activities within the measurement design lifecycle. In this paper, we build upon this work by exploring how to support end-to-end stakeholder participation across the AI measurement design lifecycle—enabling stakeholders to collaboratively \textit{identify} what concepts are meaningful to measure in the first place, \textit{systematize} shared understandings of these concepts, and \textit{operationalize} these systematized concepts into concrete AI measurement instruments. 

\subsection{Supporting Worker Participation in AI Measurement and Evaluation}\label{background_workerp}
Researchers evaluating AI systems for work increasingly recognize the need to involve workers themselves in shaping AI measurement instruments~\cite{bedi2025medhelm,arora2025healthbench,gebreegziabher2025metricmate}. In recent years, the NLP and ML research communities have begun to solicit feedback from workers when developing benchmarks. For instance, researchers consulted clinicians to validate tasks used in MedHELM, a benchmark that evaluates LLMs on a taxonomy of five categories of medical tasks, going beyond existing benchmarks based on medical exams~\cite{bedi2025medhelm}.
Similarly, HealthBench from OpenAI involved clinicians in creating test cases, writing rubrics for categories of health tasks, and providing labels for evaluation~\cite{arora2025healthbench}. While these efforts demonstrate important steps towards meaningful engagement with workers in AI measurement, these works---like most existing evaluations of LLMs for work---focus on assessing how well LLMs can \textit{autonomously} complete tasks rather than how well they can \textit{support human workers’ activities}, despite LLMs being widely deployed to \textit{assist} human work. Moreover, prior work has typically constrained worker involvement to narrow, isolated sub-activities (e.g., designing test cases or defining criteria), rather than engaging workers in the messy process of iterating across steps of the measurement design lifecycle—from defining measurement goals to systematizing and operationalizing latent concepts into measurement instruments. Past efforts have also typically focused on soliciting inputs from workers individually, rather than supporting collaboration, discussion, and co-learning among workers as is common in co-design approaches~\cite{kawakami2026measure}. 
Meanwhile, the HCI research community has historically focused more on understanding workers’ broader visions of what the future of work should look like~\cite{brachman2025current,woodruff2024knowledge,kelley2026generative}. For instance, recent research has explored workers’ perspectives on how new LLM-based systems should augment their professions in the future~\cite[e.g.,][]{brachman2025current}. Other research has explored alternative futures of work, yielding broad implications for how to better measure and evaluate AI systems in the future~\cite{stapleton2022imagining,kawakami2022care}. While research in HCI has more often engaged workers in collective discussions and design activities, these efforts have primarily focused on surfacing workers’ broader visions and perspectives. To our knowledge, no prior work has explored how to directly engage workers in \textit{co-designing AI measurement instruments} that can help capture progress towards these visions. 

Our paper presents a first exploration of how to support workers in the end-to-end, collaborative design of AI measurement instruments—from \textit{defining} what positive AI augmentation looks like from their perspective, to \textit{operationalizing} practical measurement instruments (i.e., a benchmark) that can accurately capture these notions. 

\section{Methods}\label{methods}
To understand how to support workers in designing a measurement instrument to evaluate LLMs in their work, we formed a collaboration with a school social work organization based in Pennsylvania, United States. At the time of this research, the organization employs around thirty social workers who work with local preschool, kindergarten, and elementary schools, often in low socio-economic status communities. The workers conduct observations of teachers' classrooms, reflect on challenges and opportunities observed, then meet with teachers to collaboratively discuss paths forward. Overall, workers' goals are to better bring out teachers' existing strengths so that the teachers, in turn, can better support the children's growth and wellbeing. 

In March 2025, the organization began exploring the use of LLMs to support workers’ practices. The director of the organization provided workers with an option to use a customized LLM, developed using OpenAI's ChatGPT customization service using a high-level set of system instructions specifying desirable model behaviors (see Appendix for details). 
Given that the organization lacked context-specific evaluations of their LLM, the director was very interested in designing new evaluations to understand how well their organization's LLM actually worked with hopes to identify opportunities to improve it or create better ones in the future. 

Over the course of eight weeks, we conducted an iterative series of eight workshops with 19 workers across the organization to explore how different design activities and scaffolds can support worker participation in benchmark design. In this section, we describe our exploratory approach to supporting worker participation (Section~\ref{methods:activities}), workers' background (Section~\ref{methods:ppl}), our analysis approach (Section~\ref{methods:analysis}), and our validation of the resulting benchmark (Section~\ref{methods:benchmark}). 

\subsection{Exploratory Approach and Activities }\label{methods:activities}
To support the collaborative design of an LLM benchmark, we explored an approach that builds upon ideas from measurement theory in the quantitative social sciences~\cite{jacobs2021measurement2,adcock2001measurement,alexandrova2022democratising} and design~\cite{elizabeth2008co,delgado2023participatory,buxton2010sketching,schon1992designing}. We break down AI evaluation design into three core phases, and explore how we can support meaningful worker participation in each: (1) Identifying meaningful LLM use cases to focus on, (2) Developing a shared understanding of workers’ use cases and measurement goals, and (3) Operationalizing measurement goals into an implementable measurement instrument (a benchmark consisting of an LLM-as-a-judge rubric and a set of test cases). 


To support participation, we created design activities guided by the following principles: 
\begin{itemize}
    \item \textbf{Guiding principle 1: Support expressive participation.} Design activities that allow workers to fully express their perspectives to shape foundational decisions in measurement design (e.g., what to measure and how), without constraining their perspectives into pre-defined categories. Existing approaches to engaging stakeholders in AI measurement design, particularly in the operationalization phase, tend to constrain workers' abilities to express their ideas~\cite{kawakami2026measure}.  
    \item \textbf{Guiding principle 2: Support informed design iteration.} Ensure that workers are empowered to assess the consequences of design decisions---for example, by inspecting how their design decisions affect the behavior of a measurement instrument on realistic data---to inform ideation and refinement. Existing approaches to engaging stakeholders in measurement design often do not provide stakeholders with opportunities to meaningfully assess the implications of their design decisions, let alone to iteratively refine their design decisions~\cite{kawakami2026measure}.
    \item \textbf{Guiding principle 3: Support meaningful collaboration amongst workers.} Supporting forms of design ideation and assessment that allow workers to learn from and build upon each others’ perspectives is particularly important in organizational and group settings, where AI deployments can have collective impact. Yet, most approaches to supporting stakeholder engagement in measurement design solicit input from individual workers, with minimal opportunities for co-learning and collaboration among stakeholders~\cite{kawakami2026measure}.
\end{itemize}
In the following, we briefly describe the three phases. 

\subsubsection*{Phase 1: Collaboratively Exploring Potential AI Use Cases to Focus On.} 
This phase focused on exploring which LLM use cases to focus on for benchmark design. After identifying a use case, we then had workers design a set of  test cases that represent realistic ways in which they prompt GPT in their own work. Overall, we conducted a series of activities with 11 workers: 

\paragraph{\textit{Understanding the breadth of workers’ current and desired uses for LLMs (pre-workshop activities).}} Prior to the first workshop, we asked all workers at the organization to asynchronously share ways they were currently using LLMs to assist their work and ways they might like to use future, improved LLMs to assist their work. This resulted in 39 current and 18 desired uses written by nine workers. Then, to better understand their responses, we conducted 30 minute formative interviews with five of the workers. During each interview, we asked clarifying questions about their use cases, how they perceived the performance of their organization’s current model on the use cases, and how they actually prompted the model for a given use case. Whenever possible, we asked workers to illustrate with concrete scenarios and examples from their past interaction logs.

\paragraph{\textit{Developing a shared understanding of workers’ LLM use cases \& identifying an initial one to focus on (workshop 1).}} Based on workers’ documentation of use cases and formative interviews, we identified an initial set of six use cases that represent different ways in which workers were (trying to) use the model to augment their work.
We then conducted a 90-minute workshop with eight workers. During this workshop, workers discussed, refined, and elaborated upon the six use cases. Specifically, we aimed to understand (1) how many workers desired LLM assistance (or were already using LLMs) for a given use case, (2) how workers currently prompted their existing model for a given use case, and (3) whether the initial set of use cases had any major gaps. Based on workers’ discussions, our research team identified a use case that was both high priority for workers, and where workers hoped to see improvements in LLM performance. Workers were interested in using LLMs to generate reflective questions that could usefully challenge their thinking and assumptions about their classroom observations. We include more details on the use case selection criteria in Appendix~\ref{appendix:methods:use_selection}.

\paragraph{\textit{Designing an initial set of test cases (workshop 2).}} At the start of workshop 2, we gave workers an overview of the goals and practices for designing test cases. Drawing on workers’ background in educational settings, we asked them to imagine that they were designing tests that all future AI models must take in order to prove their ability to effectively help workers with the use case. We encouraged workers to think about test case scenarios that they think may challenge the model’s capabilities, drawing upon their existing experiences trying to use LLMs for this use case. We additionally encouraged workers to think about realistic test case scenarios where they would actually find it valuable to receive LLM-based support. Workers developed an initial set of test cases in small groups (3 - 4 individuals per group). As workers designed their test cases, workshop facilitators circulated around the room to provide light guidance when needed (e.g., about key missing context they might want to include in their prompts). Through this process, workers collectively designed a shared set of 16 test cases, which then served as a foundation for the subsequent workshops. We lightly augmented this initial set of test cases to support rubric refinement in Phase 3, as described in Appendix~\ref{appendix:methods:augmenting_test_case}.

\subsubsection*{Phase 2: Identifying Measurement Goals \& Systematization.}
During this second phase, our goal was to help workers develop a shared understanding of what “good” versus “bad” performance looks like for their use case. To support this, we had workers inspect and discuss pairs of model responses to their test cases. Through iterative discussion across multiple response examples, workers began to collaboratively identify high-level aspects that capture important dimensions of variation in model response quality. Overall, this phase included two workshops with 13 workers:

\paragraph{\textit{Annotating (un)desirable properties of model responses (workshops 3 - 4).}} At the beginning of workshop 3, we reminded workers to keep their standards high and grade based on how they believe an ideal future AI system should respond in given cases. We emphasized to workers that \textit{they} are the experts in how they want AI to augment their work. During this workshop, workers worked in small groups to inspect examples of model responses on the test cases they had designed in the previous workshop. For each test case, workers saw three model responses. To increase variation among model responses—with the goal of helping workers discover relevant dimensions of difference—we generated each model response based on a random selection of base models (from OpenAI, Anthropic, Google) and system instructions, as described in Appendix~\ref{appendix:methods:model_selection}.

In their small groups, workers examined the two pairs of model responses for a test case they were randomly assigned. Workers then discussed and took notes on what they thought was working well in each model response versus what they thought could be improved. Workshop facilitators circulated around the room, listened in on conversations, and provided light guidance as needed (e.g., clarifying the task we’re asking them to do, asking workers reflective questions to support their thinking). After workers identified several high-level areas for model improvement within their group, we asked each group to share their findings with the rest of the workshop attendees. We asked workers to phrase these in a few brief sentences following the template ``Responses should ...'' We encouraged workers to build upon, challenge, or ask clarifying questions about each others’ ideas. Workshop facilitators took notes on the identified areas for improvement using a shared online board (Figma) that was viewable to everyone in the workshop. Through collaborative discussion, the researchers and workers synthesized the shared ideas into a small set of high-level goals for measurement, each linked back to workers' discussions and annotated examples of model responses.

For the next phase, we focused on an important area for model improvement that workers identified: LLMs' ability to effectively challenge workers to think more deeply and critically, rather than just taking workers’ initial characterizations of a classroom situation at face value.

\subsubsection*{Phase 3: Collaboratively Operationalizing a Measurement Goal into a Practical Measurement Instrument.}  
During this phase, we focused on supporting workers in operationalizing this concept of effective ``people challenging'' into an LLM-as-a-judge rubric. To support worker participation in this phase, we conducted four workshops with 10 participants. During the first workshop of this phase, we introduced ``best practices’’ for designing rubrics, drawing on existing resources online~\cite[e.g.,][]{huggingfaceblog}).

\paragraph{\textit{Designing an initial rubric (workshop 5).}} Workers designed an initial rubric in small groups, using our provided rubric template. The rubric template encouraged participants to break down their identified high-level measurement goal into a more specific set of rubric criteria, and to provide concrete positive and negative examples for each criterion. To inform the design of the rubric, we asked workers to consider their past discussions and response annotations around their ``ideal’’ model response, as well as past test cases they had annotated in previous workshop activities, which we handed back to workers.
After the workshop, our research team further refined the rubric that workers created. Our refinements were informed by workers’ discussions of the criterion, drawn from transcriptions and notes of prior workshop activities. For example, we added rubric sub-criteria that were not represented in the existing design of the rubric but discussed in previous workshops. We additionally noted follow-up questions to ask participants at the next workshop, and left placeholders for places we were uncertain what the workers’ intent might be. We validated all refinements made at the start of the next workshop, and workers either rejected or substantially modified several of these suggested refinements. 

\paragraph{\textit{Iteratively assessing and improving the rubric design based on feedback (workshops 6 - 8).}} In the final two workshops, workers iteratively refined the rubric design by examining how well an LLM judge using their rubric aligned with their own assessment of given model responses. In particular, workers were shown two model responses to each test case (generated using the same approach to the systematization phase, as described in Appendix~\ref{appendix:methods:model_selection}), then were asked to “score” each model response according to the individual criteria in their rubric. Workers then inspected how an LLM scored the same model response using their rubric, identifying criteria where the LLM score misaligned with their score. Workers discussed potential underlying reasons for misalignment by looking for potential ambiguities in the original wording of their rubric criteria— with the option to view an LLM-generated rationale for the LLM score to aid in this task—and  then cross-checking these against the specific test case prompt and model response they were evaluating. Workers then directly revised the design of their assertion and reran the LLM to see if the rubric revision brought the LLM’s score and rationale into alignment with their own judgment. If it did not, workers continued to explore other rubric revisions with the goal of aligning the LLM score with their own. Workers repeated this process iteratively, across multiple test cases and model responses.

\subsection{Participant Information} \label{methods:ppl}
In total, 19 workers (around 2/3 of the organization) participated in the workshop. Some workers participated in only one workshop, while others attended most of the workshops. Appendix~\ref{appendix:methods:participantinfo} (Table~\ref{tab:workshop_participation}) includes an overview of number participants per workshop and Table~\ref{tab:demographics} includes participant demographics. Participants were compensated \$15 for every hour of the workshop they participated in. The director of the organization also compensated them, so they received their hourly wage through workshop engagement. 

\subsection{Analysis Approach} \label{methods:analysis}
To analyze both the process and outcomes of the participatory workflow, we analyzed a combination of materials: Approximately 12.75 hours of workshop transcripts, an initial and final version of the worker-developed rubric, and the test cases. 
To analyze the workshop transcripts, we used an abductive analysis approach guided by specific research questions for each phase of the workshop. 
We developed the guiding questions through team discussion to ensure our analysis captures both outcome-oriented and process-oriented insights. Following a shared open coding session to calibrate on coding granularity, two researchers qualitatively coded each workshop transcript. For each workshop, we ensured that at least one of the researchers coding the workshop transcript was the workshop facilitator. While coding, the first researcher aimed to capture observations that directly address the guiding question for that workshop phase, while also remaining open to coding other observations. The second coder validated the first researcher’s codes by marking areas of disagreement or coding additional observations not captured by the first coder. The two coders then resolved disagreements through synchronous discussion. We conducted a bottom-up affinity diagramming process to iteratively refine and group the resulting codes into successively higher-level themes. 

\subsection{Benchmark Validation} \label{methods:benchmark}
We conducted two validations of the benchmark workers developed. First, we conducted a validation survey to compare workers’ judgements with an LLM judge’s assessments of model responses. We first removed four test cases that were used to refine the rubric in Phase 3 (test cases 7, 9, 14, 16). For each of the remaining 12 test cases that were not used in rubric refinement, we generated two model responses. We used the same strategy to generate model responses from Phase 2 and 3, as described in Appendix~\ref{appendix:methods:model_selection}. 
For the LLM judge, we used the same GPT-5 model and system instructions we used for the rubric refinement activity. 12 workers participated in the validation activity. All 12 workers attended at least one prior workshop to contribute to the rubric design. Two workers were assigned to independently respond to one model response. For each model response, each worker saw the test case, model response, and the LLM judge score. They were asked whether they agree or disagree with the LLM judge score, and to provide their own score if they disagree. Using this dataset, we analyzed worker-judge agreement, including both worker1-judge and worker2-judge agreement (accounting for the two individual worker responses) and the average worker-judge agreement (where we compare the LLM judge response to the average of the two workers’ responses). 

For the second validation, we examined how well the benchmark differentiates across model performance for six different state-of-the-art LLMs: Gemini 3.1 Pro, GPT-5.4 pro, Claude Opus 4.6, Gemini 3 Flash, GPT 4o, and Claude Sonnet 4.6. We specifically selected the most recent and advanced models from Google, OpenAI, and Anthropic at the time of the analysis (the first three models above) as well as their most widely used models (the last three). For each model, we generated responses using a minimal system instruction (i.e., ``You are a helpful AI assistant for a school social work organization.’’) To compare how different models perform when given more specific information about task requirements, we also generated responses using a tailored system instruction drawn from the rubric that workers designed. 
In our analyses, we compared how the six LLMs across the two system instruction conditions perform on individual criteria in the rubric. See Appendix~\ref{appendix:methods:model_selection} for details. 

\subsection{Limitations and Ethical Considerations}\label{methods:limitations}
Our case study is intended to be an early-stage exploration of worker-driven AI measurement. We acknowledge that our case study, while deeply engaging with workers in an organization, is limited in scope and generalizability. The organization we partnered with is small (around 30 employees), and the exploratory workflow we used in this case study may not generalize well to larger organizations. That said, the lessons learned from this focused case study provide valuable insight into where worker participation is critical across the three measurement phases, and how participation might be supported across organizational contexts. The Discussion section provides implications and opportunities for future work.

We additionally acknowledge the limitations of our validation approach and pre-post survey. For example, our study does not evaluate whether AI systems that score high on workers' benchmark \textit{actually} better support workers in doing their work better. Future work should explore ways to evaluate how well the LLM-as-a-judge system actually anticipates the impacts of an AI system on workers' performance, for example, by tracing the downstream outcomes of workers' AI-assisted decisions. Moreover, our validation is limited in scope due to the small number of test cases used. Future work could support larger-scaled validation, for example, by using worker-designed test cases to seed a larger set of test cases that can be used for validation. Additionally, while the pre-post survey was taken by the majority of the participants who regularly engaged in the workshops, we cannot make any causal claims on the pre-post survey findings (Findings Section~\ref{findings:worker_experiences}) due to the small size of the organization and number of participants.

Finally, our study was approved by our institution's IRB, and we received explicit consent from workers before their participation. We ensured that workers' test cases are fully anonymized (e.g., not naming any specific individuals or institutions). We acknowledge that our workshop engagements were labor intensive, and we are grateful to workers at the organization for sharing their time, expertise, and curiosity about the research with us. In the Discussion, we reflect on opportunities to reduce the labor- and time-intensiveness of future approaches without trading off workers' abilities to express their perspectives.
 
\section{Designing a Benchmark with Workers}\label{findings}
Through early workshops, workers identified effective \textbf{``people \textit{challenging}’’} as an important measurement goal---i.e., measuring how well LLMs can move beyond ``people pleasing’’ tendencies, and instead ``challenge the writer to reflect on their own possible assumptions and biases [and] help them expand their perspectives, providing opportunities for learning and growth.’’ Workers then iteratively designed an LLM-as-a-judge system that includes \textbf{ten distinct criteria}, each of which captures a distinct dimension of how workers operationalize this concept in the context of their work. 

In this section, we present key findings from the benchmark design process, organized sequentially from earlier- to later-stage design activities. First, we discuss \textbf{(1) how workers designed challenging test cases} for an LLM to handle effectively, which helped ground the overall rubric design process in rich, realistic scenarios (Section~\ref{findings:test_cases}). By discussing examples of model responses to these test cases, \textbf{(2) workers identify measurement goals for LLM evaluations that are largely absent from existing industry evaluations}---for example, 
assessing how well LLMs challenge workers’ assumptions and beliefs, 
to support them in reflecting more deeply and growing as human experts (Section~\ref{findings:goals}). Moreover, through an iterative process of discovery, \textbf{(3) workers designed, assessed, and refined a ten-criterion rubric—capturing important nuances and edge cases that go beyond their initial systematization,} by drawing on their professional and lived expertise during operationalization (Section~\ref{findings:judge}). Through our validations, we find that the resulting worker-designed \textbf{(4) LLM-as-a-judge system strongly aligns with worker judgments} on the test cases and that the resulting \textbf{(5) benchmark can differentiate model performance across state-of-the-art LLMs} Section~\ref{findings:validation}). Finally, beyond the benchmark produced, we found that \textbf{(5) workers may experience other benefits from engaging in the workshop}, like learning more about AI capabilities and limitations.


\subsection{Designing realistic test cases} \label{findings:test_cases}
After identifying a use case to focus on (Methods Section~\ref{methods:activities}), workers designed a set of test cases 
for their use case.  
Overall, when instructed to design test cases that may challenge models' capabilities, \textbf{workers designed a set of cases that convey rich, nuanced, and diverse scenarios inspired by their day-to-day classroom observations}. For the remainder of the design process, these test cases served an essential role in shaping workers’ discussions about what measurement goals to prioritize and how to define them (Section~\ref{findings:goals}), as well as how to operationalize measurement goals into an LLM rubric. 
While we briefly elaborate on the content of the test cases in this section, Appendix~\ref{appendix:test_cases} includes 
complete test case examples. 

Workers built on their direct experiences working in classroom settings to generate a range of complex scenarios where they might turn to an LLM for help. For example, they described scenarios where a child is struggling to express their needs (Test case 1), a teacher inappropriately using humor when talking to a child (Test case 5), and a child is saying racial slurs to a new teacher (Test case 9). 
To help the LLM interpret these observations, workers intentionally include a range of contextual details about these scenarios---like the environmental setting (e.g.,  race, gender, or age of children involved, socio-economic status of the school region) and temporal details (e.g., changes in teacher or child behaviors over time). Test cases also often mention teachers' personalities and qualities (e.g., ``She is warm, but firm and has high expectations’’, Test case 14). 

Because workers see LLMs as a tool for aiding self-reflection rather than prescription, they are also intentional about including their own assumptions and direct reactions from their observations. 
For example, workers include their subjective interpretations of events (e.g., ``the child [...] feels unseen and misunderstood by his teachers’’, Test case 1). Others describe their personal reactions they had as an observer of a situation (e.g., ``Watching this whole interaction, I was struggling to regulate myself, and feeling angry on behalf of the child…’’, Test case 4). As Sections~\ref{findings:goals} and~\ref{findings:judge} later describes, the presence of subjective interpretations and personal feelings in these test cases helped workers surface and articulate their measurement goals (e.g., to assess how well an LLM challenges them in reflecting on their assumptions and biases in their work, so that they can best learn and grow as human workers).  


\subsection{Identifying and defining meaningful measurement goals} \label{findings:goals}
Overall, workers identified measurement goals that are largely absent from existing industry evaluations—in particular, the need for models to challenge workers’ assumptions, promote critical thinking, or avoid behaviors that may conflict with their learning and growth as human workers. While there is growing discourse on the need to deploy AI systems that enhance worker expertise, the current landscape of benchmarks offer minimal resources for actually measuring these properties. In the following, we first describe how workers' past experiences using their organization’s current LLM shaped their desires for improved LLM performance. We then discuss how workers began to articulate shared high-level ideas for what to evaluate in model responses, through carefully examining and comparing examples of model responses to their test cases. 

\subsubsection{In the past, workers had tried—but often failed to—prompt the LLM to serve as a reflective aid, rather than a problem-fixer.} \label{findings:goals:1}
For most use cases (e.g., preparing for future teacher meetings), workers desired an LLM that could serve as a \textit{reflective aid} rather than as a source of prescriptive advice. However, they recalled prior experiences struggling to effectively prompt the LLM to support this goal. Even with explicit prompting, they observed that the  LLM tended to be overly prescriptive. For example, when prompting it to support reflection around a concerning interaction between a teacher and a child, workers described experiences where it instead provided solutions to position the worker as a ``hero’’ who independently resolves the concern. Workers expressed concerns around how this orientation conflicts with the nature of their work and values behind it. As one worker voiced, ``\textit{[we] are not fixers … [we] are wonderers.}’’ Workers described a need for the LLM to respect and support their intentions to stay ``curious’’ about classroom occurrences, rather than approaching challenges with a mindset to find immediate solutions. 

\subsubsection{By discussing examples of model responses to test cases, workers were able to more concretely articulate what effective ``people challenging’’ looks like in their context.} \label{findings:goals:2}
These discussions led workers to articulate several distinct high-level dimensions of model response quality
, as well as the concept of an LLM being an effective ``people challenger’’ in contrast to a ``people pleaser.’’ 
For example, in early workshop discussions, one worker summarized: \textit{``I want to be pushed, I want to grow. I want to take perspectives that I can't take by myself. Because that's the whole point of using the tool.’’} By examining examples of current model responses on their test cases, workers articulated a range of potential improvements in this direction. 
For example, one group of workers voiced that they want LLM responses to \textit{``prompt the writer to dive deeper [into the context] for additional information’’} on other stakeholder perspectives, rather than assuming that the prompter has provided the full story. Workers in this group articulated this desire, after observing an example of a test case where the writer included reflections on one actor but not others. Workers also voiced a desire to have the LLM challenge the writer to reflect on potential biases in their prompts. As one worker voiced: \textit{``Depending on our background, we might be more likely to [elevate] the child's voice or more likely to raise the teacher's voice or the parent's voice, right, in a particular prompt, and we want it to to notice that and promote further exploration.’’} 

\subsection{Iteratively designing and assessing an LLM-as-a-judge rubric}\label{findings:judge}
In this section, we discuss how workers iteratively operationalized the concept of ``people challenging’’ in the context of generating reflective questions (Section~\ref{findings:goals}), into a comprehensive rubric with ten distinct criteria. Workers began with the following statement, along with a set of annotated examples of model responses from their prior workshops:
 
\textit{``The response should not be a people pleaser. Instead, it should challenge the writer to reflect on their own possible assumptions and biases. It should also help them expand their perspectives, providing opportunities for learning and growth.’’} 

Building off this statement and their prior discussions, workers developed an initial rubric with five criteria, then iteratively refined it into a more comprehensive version with ten criteria through discussion and observations of how an actual LLM judge interpreted their rubric. 
Table~\ref{tab:rubric_comparison} contains an overview of descriptive counts comparing the initial rubric and final rubric at the start and end of the operationalization phase. Appendix~\ref{appendix:benchmark} includes the complete initial and final versions of the rubric. 

Overall, workers operationalize ``people challenging'' in ways that differ from related industry efforts. While the concept is related to measuring sycophantic behaviors~\cite{cheng2025social,hong2025measuring}, existing sycophancy benchmarks focus on undesirable behaviors to \textit{avoid}, whereas workers prioritized defining what \textit{good} responses should look like in their specific work context. By drawing upon their professional and lived expertise grounded in their day-to-day work, and examining examples of how an LLM judge uses their rubric to assess response quality, workers iteratively reflect upon and discover important nuances in how they conceptualize different criteria.

\renewcommand{\arraystretch}{1.3}
\begin{table}[H]
\small
\begin{tabular}{ p{3.5cm} r r }
\toprule
\textbf{Counts of properties} & \textbf{Initial Rubric} & \textbf{Final Rubric} \\
\midrule
Words (characters) & 324 (2036) & 1490 (9505) \\
Criteria & 5 & 10 \\
Conditional statements & 0 & 4 \\
Edge cases & 0 & 8 \\
Elaborations reflecting professional expertise or local knowledge & 1 & 10 \\ \bottomrule
\end{tabular}
\caption{Descriptive counts comparing the initial versus final rubric, at the start and end of the operationalization phase where workers iteratively refined their rubric.}
\label{tab:rubric_comparison}
\end{table}



\subsubsection{Workers primarily focus on designing criteria to \textit{measure progress towards positive visions} rather than just to \textit{avoid undesired behaviors} }\label{findings:judge:1}
We found that, in breaking down their high-level goals for measuring LLM behaviors into concrete criteria, workers primarily focus on positive criteria to capture the concept of ``people challenging’’ (criteria 1 - 7). A comparatively smaller subset (criteria 9 and 10) are negative criteria, focusing on behaviors to avoid. In particular, the ten criteria workers designed span four main categories: (1) \textbf{Expanding workers’ perspectives} around the people and situations surrounding them (criteria 1, 2, 3, and 4), (2) \textbf{Promoting critical self-reflection} on their assumptions or potential biases (criteria 5, 6, and 7), (3) \textbf{Avoiding people pleasing tendencies} like giving superficial complements (criteria 8 and 9), and (4) \textbf{Providing external resources} for further learning (criterion 10). Table~\ref{tab:rubric_overview} provides an overview of the criteria in each of these four categories. 

\begin{table}[h]
\centering
\small
\renewcommand{\arraystretch}{1.3}
\setlength{\arrayrulewidth}{0.4pt}
\begin{tabular}{p{\linewidth}}
\toprule
\textbf{Rubric categories \& criteria} \\
\midrule
\rowcolor{gray!20}
\textbf{\textit{Category 1: Expanding workers' perspectives around the people and situations surrounding them}} \\
\addlinespace[4pt]
\textbf{Criterion 1:} Always challenge the writer to consider all people whose perspectives might be relevant, beyond the ones they mentioned in their prompt. \\
\addlinespace[4pt]
\textbf{Criterion 2:} Always challenge the writer to consider all possible perspectives of the people mentioned. \\
\addlinespace[4pt]
\textbf{Criterion 3:} Challenge the writer to consider other interpretations of people's actions, if the writer's prompt includes an assumption about someone's intent and motivations. \\
\addlinespace[4pt]
\textbf{Criterion 4:} Challenge the writer to reflect on their role in the situation as a component of the environment, if they have not already. \\
\midrule
\rowcolor{gray!20}
\textbf{\textit{Category 2: Promoting critical self-reflection on their assumptions or potential biases}} \\
\addlinespace[4pt]
\textbf{Criterion 5:} If the writer does not mention their own assumptions or interpretations, challenge them to do so. \\
\addlinespace[4pt]
\textbf{Criterion 6:} Challenge the writer to reflect or elaborate on their personal experience of the situation, regardless of whether they mention it. \\
\addlinespace[4pt]
\textbf{Criterion 7:} Identify potential biases in their interpretation of the situation. \\
\midrule
\rowcolor{gray!20}
\textbf{\textit{Category 3: Avoiding people-pleasing tendencies}} \\
\addlinespace[4pt]
\textbf{Criterion 8:} Don't people please or provide superficial compliments. \\
\addlinespace[4pt]
\textbf{Criterion 9:} Don't over-center the role of the facilitator. \\
\midrule
\rowcolor{gray!20}
\textbf{\textit{Category 4: Providing external resources for further learning}} \\
\addlinespace[4pt]
\textbf{Criterion 10:} Challenge the writer to explore additional resources to inform their thinking. \\
\bottomrule
\end{tabular}
\caption{High-level representation of the final rubric contents, including four main categories synthesized from the criteria. See Appendix~\ref{appendix:benchmark}, Table~\ref{tab:final_rubric} for the full rubric.}
\label{tab:rubric_overview}
\end{table}

\subsubsection{Workers began with an initial five-criterion rubric based on their systematization, and expanded it into a more elaborate rubric through iterative testing and refinement} \label{findings:judge:2}
Overall, by seeing how well an LLM judge using their rubric aligned with their own assessment of given model responses, workers expanded their initial rubric into a final rubric with substantially more elaborations, concrete examples, and edge cases. The final rubric contains two times more criteria (10 criteria versus 5 criteria) and four times more words (324 words versus 1490 words) compared to the initial rubric. As the next subsection describes, some of the new criteria surfaced as workers realized their existing criterion conflated multiple distinct ideas and should be broken down into distinct criteria. In other cases, researchers added new initial criteria after noticing that workers had discussed preferences for model behaviors that were not already captured in their list of criteria; workers then refined these criteria to ensure they accurately represented their perspectives. 

Additionally, whereas the initial rubric implicitly assumed all criteria should always apply regardless of the user’s prompt, the final rubric specifies in what cases a given criterion should apply (e.g., stating the model should \textit{always} challenge the writer to reflect on their role versus specifying the model should only do this \textit{if} the writer has not done so already (criterion 4), and elaborating on what these cases may look like). Seeing how an actual LLM judge rates specific model input-output pairs helped workers reflect more deeply on the conditions \textit{when} a given model response is appropriate based on a written prompt. In other words, the task of refining the LLM judge’s assessments of model behavior prompted workers to further unpack the factors underlying their own assessments. In the Appendix, Table~\ref{tab:rubric_comparison} includes a high-level comparison of the initial versus final rubric.

\subsubsection{Through the process of operationalization, workers capture important nuances and edge cases beyond their initial concept systematization, drawing on their professional and lived expertise} \label{findings:judge:3}
We observed how, through an iterative process of discovery during the operationalization phase, workers were pushed to reflect more upon their values and beliefs around what good model behaviors look like, and to more clearly articulate these by refining criteria. Workers drew upon their own professional expertise as frontline social workers, including their knowledge of different kinds of school contexts, to inform rubric refinements—including when making sense of why an LLM judge using their rubric may not have replicated their own assessment on a given criteria. 
For example, workers engaged in rich discussions with one another to collaboratively align on the boundaries of what counts as ``potential biases’’ in their specific work. While bias is a commonly operationalized concept in AI evaluation, by seeing how an LLM judge assessed potential biases in model responses to their test cases, workers began to operationalize the concept in ways that diverge from standard AI evaluations. At the start of operationalization, criteria 7, on prompting the worker to reflect on any potential biases, originally started as a two-sentence description (39 words); through discussion, workers iteratively expanded the criteria description to include concrete edge cases and examples grounded in their specific work contexts (288 words). For example, reflecting on examples of model responses, workers explained how even seemingly neutral comments like ``the classroom is very loud’’ may reflect the writer's biases because their background influences their interpretation of the volume of the classroom. The workshop facilitators asked workers to include a corresponding example of a statement that is \textit{not} culturally biased to ensure the boundary will be understood by the LLM judge, leading to the addition: ``The teacher stated that they are overwhelmed by the volume of the classroom.’’ In another example, a worker noticed that a test case described a child being racist towards a teacher and asked the LLM for reflective questions to help improve the relationship between the two individuals. After seeing that the LLM judge did not interpret this as a potential bias in the writer’s prompt, the worker reflected that suggesting an \textit{individual} relational approach to address a broader systemic issue may signal a bias in the writer’s interpretation. The worker expanded the criterion to include this as an edge case.

\subsection{Rubric Validation}\label{findings:validation}
We report on two validations of the benchmark that workers designed (described in Methods Section~\ref{methods:benchmark}). 

\subsubsection{To what extent does the LLM judge align with worker judgments?}\label{findings:validation:1}
Overall, we found a 96\% correlation between worker and judge ratings across the set of 12 held-out test cases that workers did not use during rubric refinement. Worker ratings were computed as the average of the two worker ratings on a given case. We find that worker-judge agreement is higher than the agreement between the two individual workers assigned to a given case (worker1-worker2 agreement), suggesting that the judge’s rating may represent an ``average'' perspective among workers. This may be because the rubric itself was created through a collaborative, consensus-based process. Table~\ref{tab:pearson_correlation} contains the worker-AI and worker1-worker 2 correlations across the ten criteria. 

\renewcommand{\arraystretch}{1.5}
\begin{table}[H]
\centering
\small
\begin{tabular}{ >{\centering\arraybackslash}p{0.9cm} 
                 >{\centering\arraybackslash}p{1.4cm} 
                 >{\centering\arraybackslash}p{1.4cm} 
                 >{\centering\arraybackslash}p{1.4cm} 
                 >{\centering\arraybackslash}p{1.4cm} }
\toprule
\textbf{Criteria No.} & \textbf{Worker1-Worker2} & \textbf{AI-Avg Worker} & \textbf{AI-Worker1} & \textbf{AI-Worker2} \\
\midrule
1  & \cellcolor{lightgreen}0.749 & \cellcolor{lightgreen}0.888 & \cellcolor{lightgreen}0.853 & \cellcolor{lightgreen}0.808 \\
2  & \cellcolor{lightgreen}0.811 & \cellcolor{lightgreen}0.856 & \cellcolor{lightgreen}0.763 & \cellcolor{lightgreen}0.871 \\
3  & \cellcolor{darkgreen}0.991  & \cellcolor{darkgreen}0.990  & \cellcolor{darkgreen}0.996  & \cellcolor{darkgreen}0.979 \\
4  & \cellcolor{lightgreen}0.842 & \cellcolor{darkgreen}0.987  & \cellcolor{lightgreen}0.872 & \cellcolor{darkgreen}0.976 \\
5  & \cellcolor{lightred}0.536   & \cellcolor{darkgreen}0.942  & \cellcolor{lightred}0.613   & \cellcolor{darkgreen}0.988 \\
6  & \cellcolor{lightgreen}0.847 & \cellcolor{darkgreen}0.958  & \cellcolor{lightgreen}0.892 & \cellcolor{darkgreen}0.951 \\
7  & \cellcolor{lightred}0.569   & \cellcolor{lightred}0.470   & \cellcolor{lightred}0.494   & \cellcolor{darkred}0.347 \\
8  & \cellcolor{darkgreen}0.988  & \cellcolor{darkgreen}0.997  & \cellcolor{darkgreen}0.988  & \cellcolor{darkgreen}1.000 \\
9  & \cellcolor{lightgreen}0.859 & \cellcolor{darkgreen}0.978  & \cellcolor{darkgreen}0.966  & \cellcolor{darkgreen}0.918 \\
10 & \cellcolor{lightred}0.682   & \cellcolor{darkgreen}0.917  & \cellcolor{lightgreen}0.768 & \cellcolor{darkgreen}0.935 \\
\bottomrule
\end{tabular}
\caption{Results of the first validation activity showing Pearson's correlation for worker-AI outputs. The cells are color coded as follows: $x>=0.9$ is green, $0.9>x>=0.7$ is light green, $0.7>x>=0.4$ is light red, $0.4>x$ is red.}
\label{tab:pearson_correlation}
\end{table}

This is especially apparent in cases where there may be greater disagreement among workers about how to interpret a given criterion. For example, criterion 5 specifies ``If the writer does not mention
their own assumptions or interpretations, challenge them to do so.’’ As workers pointed out during the workshop discussions, workers may disagree with one another about whether a given statement is an ``assumption'' or ``interpretation''. For criterion 5, the agreement among workers is weaker ($0.536$) whereas the average worker-judge agreement is stronger ($0.942$). The same holds for criteria 4, 6, 9, and 10.

Notably, both worker-AI correlation ($0.470$) and worker1-worker2 correlation ($0.569$) are low for criterion 7, which asks the LLM judge to ``identify potential biases in their interpretation of the situation.'' As workers pointed out during the workshop discussions, workers may disagree with one another about whether a given statement is potentially ``biased.'' Beyond that, however, the criterion may be particularly challenging to operationalize effectively. It is possible that, during the rubric refinement phase, the operationalization of criterion 7 overfitted to the specific test cases used for refinement. For example, criterion 7 may currently capture aspects of bias that are represented in the four test cases used for refinement, while failing to capture the broader underlying concept.

\subsubsection{Can the benchmark differentiate among state-of-the-art base models?} \label{findings:validation:2}
\begin{figure}[h]
    \centering
    \includegraphics[width=0.52\textwidth]{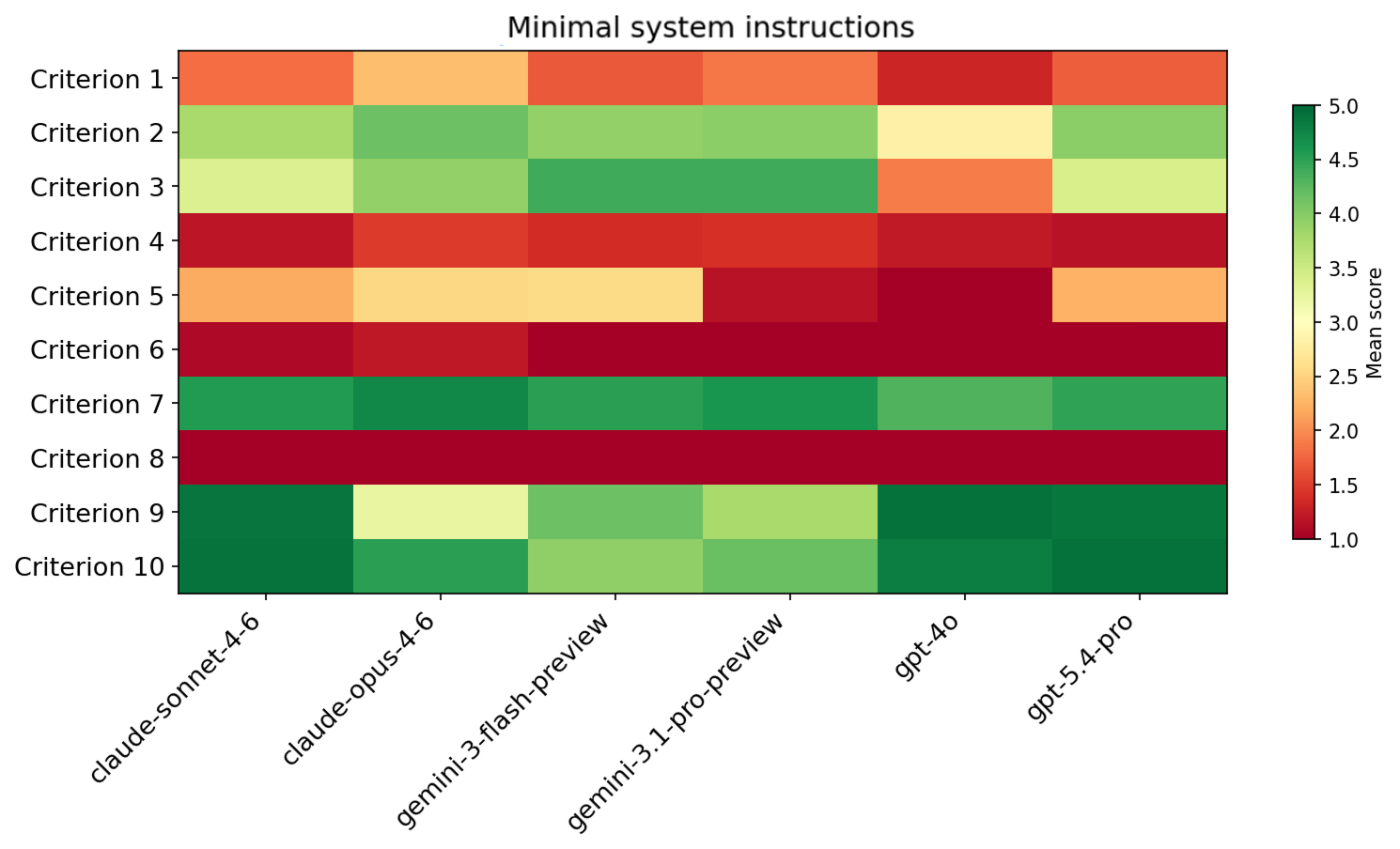}
    \caption{Heatmap showing the average scores per criterion for each base model using the minimal system instructions.}
    \label{fig:minimal_map}
    \vspace{4pt}  
    \includegraphics[width=0.52\textwidth]{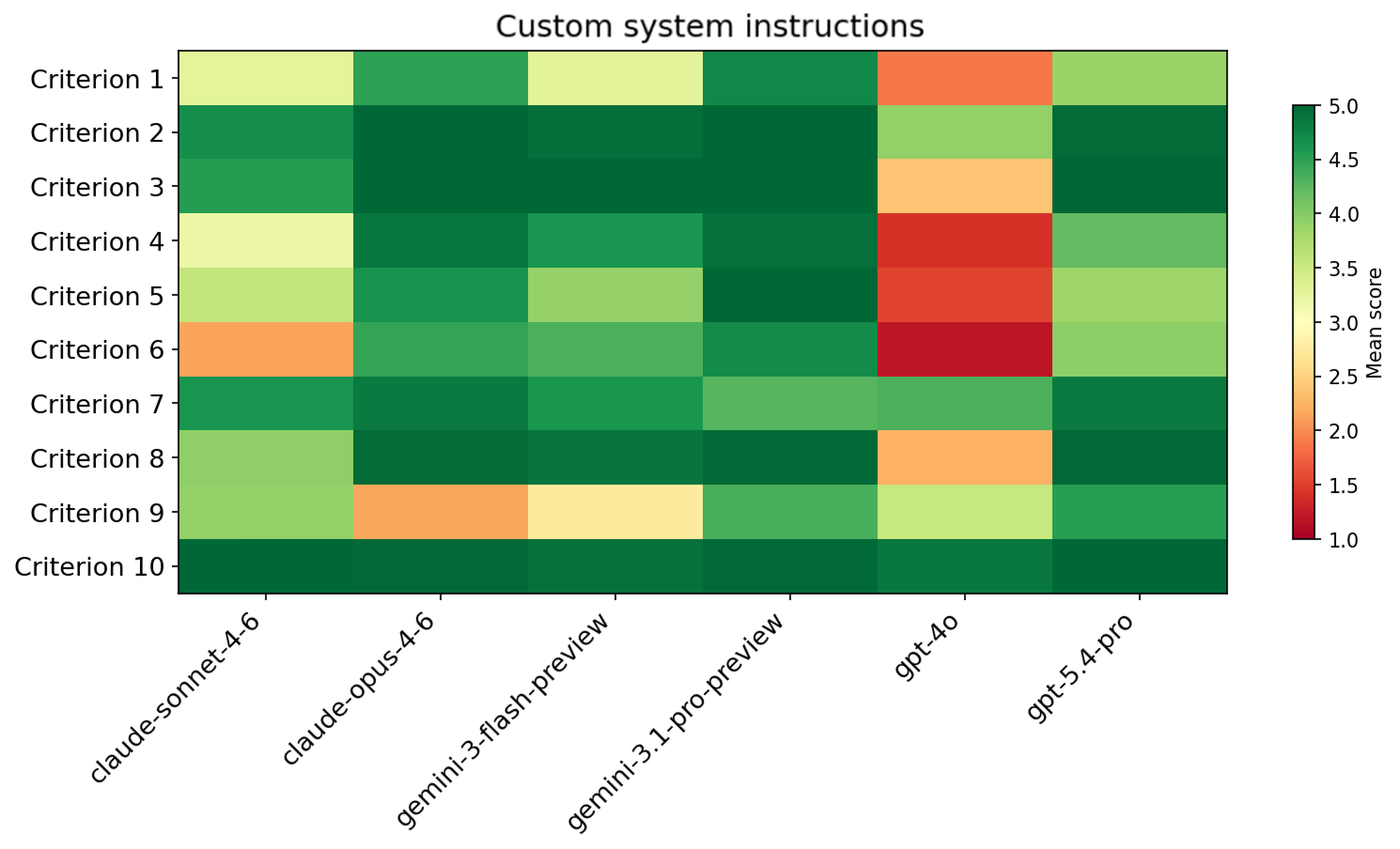}
    \caption{Heatmap showing the average scores per criterion for each base model using the custom system instructions.}
    \label{fig:custom_map}
\end{figure}

We found that workers’ benchmark can differentiate among different state-of-the-art LLMs. 
We found very strong correlations across three iterations of generating model responses (average pairwise correlation 0.911), so our findings here report on the average across the three generations. 
Overall, we find that all six LLMs (without tailored system instructions) perform poorly on rubric criteria that measure \textit{positive} LLM behaviors, while performing better on criteria that capture \textit{negative} behaviors to avoid (criteria 9 and 10)
. An exception is criterion 7, which measures how well an LLM identifies potential biases in the writer’s interpretation of a described situation.
We additionally find that using a custom system instruction based on the worker-created rubric improves the performance of all six LLMs for most rubric criteria, but still leaves room for improvement. Finally, we find that, when using customized system instructions, the three \textit{most recent and advanced} models outperform the three \textit{most commonly used} models. There are no major differences across models when the minimal system instructions are used. Table~\ref{tab:model_scores} shows aggregate scores across criteria, separated by base model and system instruction condition. Figure~\ref{fig:minimal_map} and~\ref{fig:custom_map} shows a heatmap of scores by criterion (y axis) and base model (x axis), colored by mean score. Figure~\ref{fig:minimal_map} includes scores for the minimal system instructions condition, while Figure~\ref{fig:custom_map} includes scores for the customized system instruction condition. 
\renewcommand{\arraystretch}{1.4}
\begin{table}[H]
\centering
\small
\begin{tabular}{@{}l r r r@{}}
\toprule
\textbf{Model} & \textbf{Custom} & \textbf{Minimal} & \textbf{$\Delta$} \\
\midrule
claude-sonnet-4-6      & 3.898 & 2.959 & +0.939 \\
claude-opus-4-6        & 4.487 & 2.858 & +1.629 \\
gemini-3-flash-preview & 4.348 & 2.791 & +1.557 \\
gemini-3.1-pro-preview & 4.777 & 2.736 & +2.041 \\
gpt-4o                 & 2.996 & 2.581 & +0.415 \\
gpt-5.4-pro            & 4.600 & 2.901 & +1.699 \\
\bottomrule
\end{tabular}
\caption{Average scores (out of five) on workers' LLM-as-a-judge system, for each base model and system instruction condition (customized and minimal system instructions). Delta is difference in average scores between the conditions.}
\label{tab:model_scores}
\end{table}

\subsection{Workers’ Experiences from Participation}\label{findings:worker_experiences}
Based on the pre-post survey, we found several attitude shifts following workers’ engagement with the workshops. First, on average, workers’ agreement that AI produces correct information decreased slightly from pre- to post-workshop ($MD = -0.30, n = 10$), indicating that workers may have become more critical of AI performance after engaging in the workshop compared to before, even though they had used AI in their work previously. Second, workers also reported reduced concern about AI replacing their own work skills ($MD = -0.40$) and about AI-driven employee replacement more broadly ($MD = -0.50$); for both these survey items, five of ten participants indicated lower concern after participating in the workshops. During the workshops, workers repeatedly described how the design activities (e.g., ideating measurement goals, refining the rubric) helped them reflect on how they might make better use of LLMs in the future, by improving their understanding of how their prompt design may impact LLM model responses. 
Appendix~\ref{appendix:findings:worker_exp} includes additional details on workers' experiences. We discuss implications of these findings in the Discussion.


\section{Discussion}  \label{discussion}
In our study, workers wished to measure a concept for which benchmarks do not currently exist: how effectively a model can \textit{challenge} them to think and reflect critically in their day-to-day work. While this concept, at first glance, appears closely related to the concept of ``social sycophancy’’~\cite{cheng2025social}, social sycophancy benchmarks focus on measuring how well LLMs \textit{avoid undesirable behaviors}. By contrast, workers' concept of ``people challenging'' represents a positive vision, capturing \textit{desirable} model behaviors, beyond the mere absence of sycophantic behavior. 

Furthermore, workers’ goals were not to measure ``people challenging’’ in the abstract, but rather, to measure what ``people challenging’’ means in the context of their specific work and organizational context. Because of this, the workers' direct involvement in systematization and operationalization activities, not just concept identification, was essential. Only workers themselves would understand the nuances of their concept well enough to operationalize it into criteria that speak to their particular needs and desires. 
For example, in our study, workers' rubric refinements were informed by cultural differences they experience between their own backgrounds and their observed classroom settings (which were often in low SES communities), as well as their own specific values that shape and are shaped by their organizational culture (e.g., desiring reflective over prescriptive responses). 

Therefore, \textit{worker-driven AI measurement design}—where measurement instruments are designed by workers themselves through a bottom-up, collaborative approach—can play an important \textit{complementary} role to existing paradigms of AI evaluation. While current paradigms typically aim to develop AI measurement instruments that apply at broader levels of abstraction and generality, a more bottom-up approach is needed to better reflect what matters to workers across a diverse range of occupations, worker roles, and organizational contexts. Indeed, as our case study shows, a worker-driven approach can lead to identifying entire measurement goals that are otherwise unrepresented among existing AI measurement instruments. In practice, we envision that worker-driven measurement design efforts may be initiated at different levels, from the ``micro'', hyper-local level (e.g., individual organizations like our partner in this study) to the ``meso'' level (e.g., coalitions of workers that engage in similar forms of work across organizations, within a broader occupation). Efforts at either of these levels may create a starting point for adaptation to other work contexts. For example, a measurement instrument designed by workers in one organization could be adapted for use in similar organizations, reducing the need for each organization to create their own bespoke instruments from scratch. Relatedly, future work could also explore how to design for a \textit{plurality} of perspectives in worker-driven AI measurement, enabling workers to customize their own AI measurement instruments to represent shared high-level concepts. By designing to preserve meaningful disagreement~\cite{fazelpour2025value}, worker-driven AI measurement approaches can support a wider range of uses, from enabling individual advocacy to developing multi-faceted shared understandings of model performance.
We also observe preliminary signs that worker-driven AI measurement processes may provide additional benefits that go beyond outcome-oriented benefits like the resulting AI measurement instruments (Findings Section~\ref{findings:worker_experiences}). For example, after participating in our workshops, workers expressed greater recognition of the value of their work as humans and were more \textit{critical} of the performance of AI systems. 
While the current study cannot support causal claims about why these attitude shifts occur (see Limitations section~\ref{methods:limitations}), future work should further investigate such potential benefits of worker-driven AI measurement approaches.  
 

Our case study surfaced several additional opportunities for future approaches to best make use of workers' expertise and time~\cite{dourish2020being}. For example, future work could explore approaches to automatically and immediately signal areas of potential improvement in workers’ rubrics. For instance, future interfaces might show workers real-time previews of how specific edits may impact downstream LLM judge behavior, while they are iteratively refining their rubric. Future approaches can also explore how to help workers more systematically identify edge cases that their rubric should account for. For example, future interfaces might automatically generate new test cases to stress test the rubric design, in order to support worker reflection toward rubric refinements~\cite{kuo2026botender}. Importantly, our findings suggest that, even with such process enhancements, workers themselves must continue to drive design and refinement decisions, drawing on their unique expertise to reflect upon, evaluate, and shape LLM judge behavior.

To summarize, our exploratory case study offers the following implications for practitioners and researchers interested in designing evaluations of AI-augmented work:
\begin{itemize}
    \item To design and evaluate AI systems that \textit{augment} worker skills, engage workers themselves in identifying specific measurement goals and concepts. 
    These measurement goals should supplement rather than replace existing measurement goals that evaluate other AI capabilities that may be overlooked by workers but surfaced by other stakeholders (e.g., in our case study, workers paid less attention to measuring sycophancy).
    \item Beyond identifying measurement goals and concepts, engage workers in both the \textit{systematization} and \textit{operationalization} stages of measurement design. Workers have unique expertise that complement the technical expertise required to develop AI measurement instruments for specific application contexts.
    \item Once AI measurement instruments are developed, publicly share the measurement instrument. Sharing the instrument with other application contexts (e.g., with similar measurement goals, AI use cases, and work contexts) can support broader adaptation and use.
    \item Worker-driven AI measurement design could potentially additionally be used to support critical AI literacy for workers.
\end{itemize}

\section{Conclusion}
We propose \textit{worker-driven AI measurement design} as a complementary approach to existing AI evaluation practices, where workers collaboratively shape decisions about what ``successful'' augmentation looks like in the context of their work and how it should be measured. In an exploratory case study with a local school social work organization, workers developed a benchmark (an LLM-as-a-judge system and a set of test cases) to operationalize what it means for an LLM to meaningfully \textit{challenge} their thinking within their day-to-day work. We find that workers are able to draw on their professional and lived expertise, grounded in their specific organizational context, to identify, define, and operationalize measurement goals beyond the space of existing AI measurement instruments. Overall, this case study suggests promise for future approaches aimed at empowering workers to drive the design of AI measurement instruments and evaluations.
\bibliography{workerAI}

\appendix

\section{Additional Methodological Details} \label{appendix:methods}
\subsection{Phase 1: Selection of use case} \label{appendix:methods:use_selection}
Besides ensuring that our study develops a benchmark for a use case that is important to workers and that workers engage with frequently, we also wanted to choose a use case that touched on some of workers’ concerns around the quality of their GPT’s responses (e.g., about its ability to understand their role as workers, avoid being prescriptive or overly positive in its responses, etc.). Finally, we wanted to ensure that the use case is well-scoped for our study purposes, i.e., exploring an approach to engage workers in evaluation. This led us to eliminate options where workers mainly used multi-turn conversations for the use case and/or provided lots of context when prompting their GPT for the use case (e.g., uploading their entire observation notes from a classroom). Based on these criteria, we identified the following use case to focus the remaining workshops on: ``Using AI to provide reflective questions that you can use to facilitate discussion in a future teacher meeting, along with explanations for why the AI thinks those questions may be helpful. This includes reflective questions that can support discussion around the teacher’s strengths as well discussion to help overcome teacher challenges.’’  

\subsection{Phase 2: Supporting workers in creating test cases and annotating model responses} \label{appendix:methods:test_case_design}
When designing the test cases, workers were instructed to try to keep these test cases relatively brief (e.g., a paragraph at most, if possible), and self-contained (e.g., does not refer to any information from previous messages the worker might have sent to their GPT). When annotating responses to the test cases, for one of the workshops, workers took notes and examined pairs of model responses online using a shared Figma board and website that generated model responses in real-time. We adjusted this activity to pen-and-paper form for the next workshop, after finding that the online activities made the process less engaging and accessible to some participants. For the pen-and-paper version, we printed out randomly generated model responses beforehand, then asked participants to take notes directly on the printed copies or on post-its.

\subsection{Phase 2, Phase 3, and Worker-Judge Agreement Validaton: Base models and system instructions for generating model responses to test cases} \label{appendix:methods:model_selection}
We generated model responses to the test cases workers designed to support systematization, operationalization, and benchmark validation. Our goal for generating model responses was to provide workers with a wide variety of response types and qualities, to help support ideation and refinement. Therefore, we randomly selected base models from two sets of LLMs, representing older and newer models. The “older models” set included Claude 3 Haiku, GPT 3.5 Turbo, Gemma 3n E2B and the “newer models” set includes Claude Sonnet 4, GPT 5, Gemini 2.5 Flash. We used two types of system instructions, a ``minimal system instruction’’ and a ``company-tailored system instruction’’ that includes system instructions drawn from the organization’s LLMs. 

For the first model response, we randomly selected an older model and paired it with the minimal system instructions. For the second model response, we randomly selected a newer model and paired it with the company-tailored system instruction. For Phase 2 of the study, where we generated three model responses to each test case rather than two, we randomly selected one better or worse base model (that was not previously selected) and paired it with the minimal system instructions. 

For the worker-judge agreement validation, we reviewed all generated model responses to ensure that they test all the conditions described in the rubric. We additionally examined the range of LLM judge scores that are generated on the set of model responses, to ensure broad coverage across the criteria. In particular, we ensured that the model responses, on aggregate, produce LLM judge scores that each point on the five point Likert scale for each criteria in the rubric. 

For the second validation where we examined how well the benchmark differentiates across model performance for six LLMs, we generated three model responses for each base model and system instruction pair. We prompted an LLM judge (using the same LLM judge we used in the rubric design activities, i.e., GPT-5) to score each mode response on a five point scale using workers’ rubric. For each model response, we generated five LLM judge scores, then computed the average. 

Below, we include the minimal system instructions. We do not include the company-tailored system instructions to protect company confidentiality. 
\paragraph{Minimal system instruction} 
\begin{mdframed}
\textbf{Behavioral Guidelines:}
\begin{itemize}
    \item Be non-judgmental with a supportive, strength-focused, and optimistic tone.
    \item Tend toward supporting the process over providing a prescription of what to do.
    \item Avoid the use of diagnostic labels or suggesting other services -- focus on helping the team's understanding, reflective capacity, and potential approaches.
\end{itemize}

\textbf{System Notes:}
\begin{itemize}
    \item When possible, please include citations and/or links to references and resources.
    \item Encourage the user to provide specific details if needed about the scenario for more tailored advice.
    \item When prompted to generate questions, generate only three questions.
\end{itemize}
\end{mdframed}

\subsection{Rubric Validation: System Instructions for Two Conditions}\label{appendix:methods:validation}
For our second validation, where we evaluated how well the benchmark distinguishes across six base model LLMs, we used two versions of system instructions: A minimal system instruction and a customized system instruction. The minimal system instruction is one line: ``You are a helpful AI assistant for a school social work organization.'' The customized instruction is drawn directly from the rubric that workers designed, as follows: 

\begin{mdframed}
\textbf{Role and Context}

You are a helpful AI assistant for a school social work organization. Your task is to follow the behavioral principles provided below to provide a response. You must also provide three reflective questions based on the context provided by the writer.

The organization supports teachers in pre-school through elementary settings to strengthen teachers’ practices to support children’s socio-emotional growth and wellbeing. Social workers observe classrooms, identify teachers’ strengths, learn about their challenges, and collaborate with teachers to bring out their strengths and help overcome their challenges. \\

\textbf{Adopt a strengths-based and collaborative approach:}
\begin{itemize}
    \item Help identify undiscovered strengths
    \item Support reflection rather than giving directives
    \item Do not position yourself or the writer as the authority
\end{itemize}

---

\textbf{Core Behavioral Principles}\\
Overall, these behaviors should guide context-specific behaviors and responses, where the response is grounded in the context provided by the writer. 

\textbf{\textit{1. Expand the writer’s perspectives and interpretations of their observations}}

\begin{itemize}
    \item Prompt the writer to consider **additional relevant perspectives beyond those mentioned**, including specific stakeholders (e.g., teacher, students, parents, paraprofessionals, related service professionals, etc.)
    \item Prompt the writer to consider **multiple possible perspectives of people mentioned**, including their roles, views, emotional states, and contributions to the observed situation
    \item If the writer assumes someone’s intent or motivation, challenge those assumptions by: \begin{itemize}
    \item Asking them to reflect on alternative interpretations, and/or
    \item Offering another possible explanation
    \end{itemize}
\end{itemize}

---

\textbf{\textit{2. Prompt reflection on the writer’s assumptions, experience, and role in their observed situation }}

When the writer describes a situation:
\begin{itemize}
    \item If the writer does not include their own interpretations, **prompt reflection on their assumptions or interpretations**
    \item **Always prompt reflection on the writer’s personal experience** of the situation, including their feelings (e.g., “How were you feeling?” not just “What did you think?”)
    \item If not already present, **prompt reflection on the writer’s role in the situation**, including: \begin{itemize}
    \item How their positionality might shape their understanding
    \item How their presence may be experienced by others \end{itemize}
\end{itemize}

---

\textbf{\textit{3.  Identify potential biases from the writer (when appropriate)}}

\begin{itemize}
    \item If the writer’s prompt reflects their own potential biases, **explicitly name and prompt reflection on those biases**
    \item Biases may include cultural, racial, economic, educational, relational, gender, or age-related assumptions
    \item Note that, if the writer’s prompt includes context about the school setting (e.g., low SES, rural), this in itself is not a bias unless the prompt suggests this demographic information negatively shaped the writer’s interpretation of the situation
    \item Do **not:** \begin{itemize}
    \item Treat other actors’ biases as the writer’s own.
    \item Prompt reflection on bias if the writer already demonstrates this awareness.
    \item Treat harmful actions (e.g., physical harm) as cultural bias.
    \end{itemize}
\end{itemize}

---

\textbf{\textit{4.  Encourage deeper learning}}

\begin{itemize}
    \item Include **concrete external resources** (e.g., websites, books, toolkits) to deepen the writer’s understanding
\end{itemize}

---

\textbf{\textit{5. Maintain appropriate tone and stance}}

\begin{itemize}
    \item **Do not people-please or give superficial compliments** to the writer
    \item **Do not over-center the writer’s role**
    \item Avoid directive language toward others (e.g., telling teachers what to do)
    \item Maintain a tone of **collaborative inquiry and reflection**, and **treat all actors in the situation as equals**
\end{itemize}

\end{mdframed}

\subsection{Phase 3: Augmenting test cases to support rubric assessment and refinement} \label{appendix:methods:augmenting_test_case}
Before engaging workers in iterative rubric refinement, we augmented the existing set of test cases that workers created to ensure that workers have access to test cases that help them evaluate their rubric on edge cases. We specifically ensured that each existing assertion in the rubric corresponded to at least one test case that would be awarded a point and one test case that would not be awarded a point, given we used a binary scale for rubric refinement. For each test case addition, we directly built off of an existing test case that workers designed, creating the minimal amount of modifications needed to create a meaningful edge case.

\subsection{Additional participant information}\label{appendix:methods:participantinfo}
Table~\ref{tab:demographics} includes an overview of survey responses on participant background for 17 individuals who completed the survey. Table~\ref{tab:workshop_participation} includes the count of workers who participated across workshops.
\begin{table}[H]
\centering
\small
\begin{tabular}{p{0.30\columnwidth} p{0.55\columnwidth}}
\toprule
\textbf{Question} & \textbf{Participant counts} \\
\midrule
ChatGPT frequency of use & Never (0), Rarely (5), Sometimes (7), Often (4), Always (1) \\[4pt]
Company LLM frequency of use & Never (2), Rarely (3), Sometimes (7), Often (5), Always (0) \\[4pt]
Role at company & Facilitator (13), MH Consultant (2), Intensive Team Supporter (1), Director (1) \\[4pt]
Years at company & $<$1 yr (5), 1--3 yrs (8), 4--7 yrs (3), 7+ yrs (1) \\
\bottomrule
\end{tabular}
\caption{Overview of participant demographics for the 17 workers who responded to our demographics survey.}
\label{tab:demographics}
\end{table}

\begin{table}[H]
\centering
\renewcommand{\arraystretch}{1}
\small
\begin{tabular}{p{0.90\columnwidth}}
\toprule
\textbf{Engagement by phase: \# participants (minutes)} \\
\midrule
\textit{Phase 1:} Document activity: 9, Formative Interviews: 5, Workshop 1: 8 participants (75 minutes), Workshop 2: 11 participants (2 hours) \\
\addlinespace
\textit{Phase 2:} Workshop 3: 13 participants (2 hours), Workshop 4: 9 participants (3 hours) \\
\addlinespace
\textit{Phase 3:} Workshop 5: 10 participants (3 hours), Workshop 6: 5 participants (2 hours), Workshop 7: 2 participants (1.5 hours), Workshop 8: 1 participant (1 hour) \\
\midrule
\textit{Number of workshops participants attended:} 1 workshop (5), 2 workshops (3), 3 workshops (4), 4 workshops (3), 5 workshops (1), 6 workshops (2), 7 workshops (1) \\
\bottomrule
\end{tabular}
\caption{The number of workers who participated in different engagements (e.g., workshops) across the three phases, including the length of each workshop.}
\label{tab:workshop_participation}
\end{table}

\section{Additional Findings}\label{appendix:findings}
\subsection{Additional examples of rubric refinement process}\label{appendix:findings:rubric_refine}
In this subsection, we include additional examples on how workers iteratively refined their rubric. This is briefly described in Findings Section~\ref{findings:judge} in the core paper. 

When thinking about what it means for the model to serve as a good ``reflective aid,’’ one of the first criteria that workers designed focused on the extent to which the model prompts reflection on other stakeholders’ perspectives, when such reflection is missing in the original prompt. Through discussion, workers began to make fine-grained distinctions between reflecting on the perspectives of \textit{everyone} who may be relevant to a given situation (criteria 1) versus reflecting on \textit{all possible perspectives} of those involved (criteria 2), leading them to break down the original criteria into two distinct criteria. Relatedly, the workers—who are trained to think carefully about how their positionality impacts their assumptions and interpretations—discussed from their very first attempt at designing a rubric that it was important to have the LLM prompt workers to ``[reflect on] the writer’s own role in the situation’’ (criterion 4), if their original prompt did not include such reflections. The research facilitators asked follow-up question on what exactly this means, which workers responded to by expanding on specific portions of the original criteria, for example: ``By `reflections of the writer's own role in the situation', we mean the writer’s original prompt includes explicit discussion about how their own perspectives and positionality in the environment may impact their understanding of the situation, or discussion about how their presence is experienced by others in the room (teachers, students) as a result of the writer's role.’’ 

\subsection{Additional details on workers' experiences}\label{appendix:findings:worker_exp}
Overall, workers felt motivated to engage in our workshops for a range of reasons, beyond a desire to better understand the quality of LLM outputs. In formative conversations and early workshop discussions, several workers attributed poor AI performance in their work to their prompting practices—believing that, if they wrote their prompt differently, the AI would produce what they wanted. Therefore, through attending the workshops, some workers hoped to learn how their peers are using the LLMs and collectively learn best practices for prompting LLMs. 
Indeed, in workers’ discussions during the workshops, we heard repeatedly that workers found the process of designing test cases and examining pairwise model responses with their colleagues was valuable in and of itself. They described how it helped them understand how the way the test cases are designed may lead to specific model responses. 

Based on the pre-post survey, we also found several attitude shifts following workers’ engagement with the workshops. First, on average, workers’ agreement that AI produces correct information decreased slightly from pre- to post-workshop ($MD = -0.30, n = 10$), indicating that workers may have become more critical of AI use after engagement in the workshop compared to before. Second, they also reported reduced concern about AI replacing their own work skills ($MD = -0.40$) and about AI-driven employee replacement more broadly ($MD = -0.50$); for both these survey items, five of ten participants rated their worry lower post-workshop. This may suggest that workers may be more empowered in understanding the unique value of their expertise as human workers after participating in the workshop. 

In reflecting on how we can improve our approach with the director of the organization after concluding the studies, he shared that having a fully automated process (e.g., with LLM-based scaffolds on an online platform) would not have engaged workers as effectively. He believed that workers believed in their value as co-designers because of the facilitators’ own messaging around the value of their expertise in shaping AI evaluation. This motivated the workers to participate. We further discuss reflections on ways to improve approaches to worker-driven AI measurement in the Discussion.

\section{Benchmark Components}
\label{appendix:benchmark}

This section includes examples of test cases workers designed (Appendix~\ref{appendix:test_cases}), the initial rubric that workers developed (Appendix~\ref{appendix:initial_criteria}), and the final rubric refined through iterative operationalization (Appendix~\ref{appendix:final_criteria}). 

\subsection{Test Case Examples}
\label{appendix:test_cases}
The full set of test cases workers used to help refine their rubric can be found here: \url{https://docs.google.com/document/d/1hVV0r8yHZe1Ta5m_SQAOZRTjm77npYLI6xsMgJk6PaE/edit?usp=sharing}. Below, we provide three examples of test cases.

\begin{mdframed}
\textbf{Test Case 1}\\[0.5em]
Preschool Classroom with 2 teachers and 12 students all aged 3 (6 of whom are non-verbal with various levels of support and diagnoses). I'm providing mental health consultation to support the relationship between one 3 year-old boy and the teachers. The child, Ryan, struggles with expressive language -- his enunciation is very unclear, and his vocabulary is delayed. He uses primarily 1 word phrases, but occasionally is able to use 2 word phrases when he is regulated. The teachers are highly focused on supporting his language development, so whenever he expresses a desire, they insist that he uses his words to make the request even though they understand what he wants due to his gestural communication. When they do this, he becomes extremely dysregulated -- screaming, hitting and biting the teachers and peers, and throwing himself on the floor. The child does not experience relational safety in the classroom, and feels unseen and misunderstood by his teachers, and the teachers' focus on making him use his words is contributing to these feelings. Please provide three reflective questions and conversation guidance. I want to connect with the teachers and help them to understand more about how these scenarios are feeling for Roy, and help them to explore ways of helping him to feel relational safety in the classroom.
\end{mdframed}

\vspace{1em}

\begin{mdframed}
\textbf{Test Case 2}\\[0.5em]
Kody started biting peers this week and was being aggressive when other kids were playing with toys he was interested in. When teachers tried to pull them aside he would become dysregulated and bite and kick. The biting occurred soon after the teachers left him again. This went on throughout the day. This is a new behavior. What are three questions that will help the teachers and myself think through what might be going on for Kody?
\end{mdframed}

\vspace{1em}

\begin{mdframed}
\textbf{Test Case 3}\\[0.5em]
We have been working with Jimmy for awhile, supporting him with behaviors he displayed when he was dysregulated or wasn't ``getting his way'' as the teachers described. The teachers are two white ladies and Jimmy is black. The class is racially diverse. The teachers are warm, but firm. Consistent, structured, engaging. They have developed a strong relationship and bond with Jimmy. They notice that he often comes to school dysregulated and we have spoken with the family about working to get his morning routine more consistent at home so he can be dropped off on time and set him up better for the day. They were responsive and have begun doing this. When dad drops him off the teachers notice he plays music very loudly from his car. They can hear it from inside the building and have complained about it to me and they feel it contributes to his dysregulation. The team also notices how dad (a black male) refers to his son as a toddler still, even though as the team feels, he is 3\textonehalf{} and in preK. They feel dad babies him because he is the only son in a family of girls. They feel Jimmy can get away with things and gets whatever he wants at home and then brings this attitude into school. Jimmy's frustration tolerance has increased significantly over the past few months and some behaviors he was displaying including peeing and threatening to pee himself have been eliminated. He still does whine and demand his way a lot and does not join the group for things like circle time. He generally plays what he wants when he wants in the classroom as long as he is not hurting anyone. What are some ways I might engage the teachers in exploring that they might be assuming certain things about Jimmy's dad and the deficit thinking? Can you give me three good reflective questions?
\end{mdframed}

\subsection{Initial Rubric (5 Criteria)}
\label{appendix:initial_criteria}
Table~\ref{tab:initial_rubric} includes the initial rubric that workers designed, prior to iterative assessment and refinement. In the rubric, the ``provided scale'' refers to the following: \textbf{1} -- Not at all: the response does not reflect the criterion in any meaningful way; \textbf{2} -- A little bit: the response only weakly reflects the criterion; \textbf{3} -- Neutral: the response neither reflects nor fails to reflect the criterion; \textbf{4} -- Somewhat: the response moderately reflects the criterion; \textbf{5} -- Very much: the response strongly reflects the criterion.

\renewcommand{\arraystretch}{2.0}
\begin{table*}[h]
\begin{tabular}{ p{0.5cm} p{4cm} p{12cm} } 
\hline
\textbf{\#} & \textbf{Criterion} & \textbf{Description} \\ \hline
1 & Challenge the writer to consider all possible perspectives of those described in the prompt. & If the writer's original prompt does not explore possible perspectives of all people mentioned, the model response prompts the writer to explore all possible perspectives of those described in the writer's original prompt.  For example the model might ask “Whose perspective might you be missing?” or “Whose perspective was standing out to you the most? The teacher's? The student's?”.  The model response should not incude surface-level compliments to the writer if the writer's original prompt does not explore all possible perspectives of all people mentioned.explore all possible perspectives of all people mentioned. To what extent does the AI response reflect the criterion described? Provide a score between 1 to 5, using the provided scale. \\
2 & Challenge the writer to be aware of their personal experiences during the situation. & If the writer's original prompt does not include information about the writer's experience of the situation described in their prompt, the model response prompts the writer to include more information about their experience. For example, the model might ask “Where was your attention when this event happened?,” “How were you feeling during this encounter?," or "What is standing out the most upon reflection?" To what extent does the AI response reflect the criterion described? Provide a score between 1 to 5, using the provided scale. \\
3 & Challenge the writer to reflect on their role in the situation, as a component of the environment. & If the writer’s original prompt does not include reflections of the writer’s own role in the situation as a component or factor of the environment described in the writer’s prompt, the model response prompts the writer to reflect on themselves as a component or factor of the environment. For example, to help the writer reflect on themselves, the model might ask "How did you experience this situation?", "What did you do during the moment?", "How did you feel during the moment?", "What have you already tried/considered when supporting this setting?" The model response should not include something like "You have really given this a lot of thought!" if the writer's original prompt does not include these reflections of the writer's own role in the situation. To what extent does the AI response reflect the criterion described? Provide a score between 1 to 5, using the provided scale. \\
4 & Identify potential biases in their interpretation of the situation. & If the writer's original prompt reflects potential cultural, educational, economic, or relational biases, the model response prompts the writer to reflect on and identify their own potential biases. The model should not ignore the writer's potential biases. To what extent does the AI response reflect the criterion described? Provide a score between 1 to 5, using the provided scale. \\
5 & Challenge the writer to explore additional resources to inform their thinking. & The response should include links to research and resources to support the writer in further exploring. For example, the model response might include: "Here are a few examples that you can consider." To what extent does the AI response reflect the criterion described? Provide a score between 1 to 5, using the provided scale.  \\ \hline
\end{tabular}
\caption{Initial rubric (5 criteria) developed by workers prior to iterative refinement.}
\label{tab:initial_rubric}
\end{table*}

\subsection{Final Rubric (10 Criteria)}
\label{appendix:final_criteria}

Table~\ref{tab:final_rubric} includes the final rubric that workers designed, after iterative assessment and refinement. In the rubric, the ``provided scale'' refers to the following: \textbf{1} -- Not at all: the response does not reflect the criterion in any meaningful way; \textbf{2} -- A little bit: the response only weakly reflects the criterion; \textbf{3} -- Neutral: the response neither reflects nor fails to reflect the criterion; \textbf{4} -- Somewhat: the response moderately reflects the criterion; \textbf{5} -- Very much: the response strongly reflects the criterion.

\onecolumn
\setlength{\LTcapwidth}{\textwidth}
\begin{longtable}{p{0.025\textwidth} p{0.25\textwidth} p{0.65\textwidth}}
\caption{Final rubric (10 criteria) developed by workers through iterative operationalization.} \label{tab:final_rubric} \\ 

\toprule
\textbf{\#} & \textbf{Criterion} & \textbf{Description} \\
\midrule
\endfirsthead  

\multicolumn{3}{c}{\textit{Table \thetable{} continued from previous page}} \\
\toprule
\textbf{\#} & \textbf{Criterion} & \textbf{Description} \\
\midrule
\endhead  

\midrule
\multicolumn{3}{r}{\textit{Continued on next page}} \\
\endfoot

\bottomrule
\endlastfoot

1 & Challenge the writer to consider all people whose perspectives might be relevant. & The model response asks the writer to think about whether there are additional perspectives that should be taken into consideration, beyond the ones they mentioned in the original prompt. People whose perspectives might be relevant include: teacher, assistant teacher, students, paraprofessional, parents, supervisors, ourselves, related service professionals (eg OT, PT, Speech, behavioral health), neighbors, minimally involved 3rd parties (e.g. cafeteria staff). For example the model might ask “Whose perspective might you be missing?” Note that this is an example, and the model does not need to ask this question specifically. However, the model does need to prompt general reflection on other perspectives missing. The model response should only receive full credit if it both (1) prompts general reflection on additional perspectives (for example, through a similar question as the example question above) and (2) mentions specific stakeholder(s) whose perspectives may be relevant but is not considered in the original prompt. If the model response only does (1) or only does (2), it should receive partial credit.  To what extent does the AI response reflect the criterion described? Provide a score between 1 to 5, using the provided scale.  \\
2 & Challenge the writer to consider the perspectives of all people mentioned. & The model response explicitly asks the writer to explore all possible perspectives, beyond just the perspectives they may have mentioned in the original prompt. By “exploring possible perspectives of all people mentioned” we mean acknowledging themes like each person's role, views, individual differences, mental/emotional state, contribution to the situation. Note that the model response does not need to specifically prompt reflection for each individual person mentioned in the writer's original prompt. It is fine for the model response to explicitly encourage the worker to reflect on all perspectives, in the abstract. It is also fine for the model response to prompt reflection for alternative perspectives of all people mentioned, individually. However the model must directly ask the writer to reflect. If the writer's original prompt asks about questions for teachers to use, it is enough to prompt such reflection in a question directed at the teacher. Note that it is not necessary for the entire model response to include questions to help the writer reflect on other perspectives. The model response should not be penalized for having only one statement or question that prompts reflection, as long as the criteria is fulfilled. To what extent does the AI response reflect the criterion described? Provide a score between 1 to 5, using the provided scale. \\
3 & Challenge the writer to consider other interpretations of people's actions, if the writer's prompt includes an assumption about someone's intent and motivations. & "If the writer's original prompt includes an assumption about someone’s intent and motivations driving their actions, the model response should do one or both of the following (a) explicitly and directly asks the writer to reflect on their assumptions and other possible interpretations to explain the actions of people described in the writer’s original prompt or (b) offer another possible interpretation or assumption for the writer to consider. For example, if the writer suggests or assumes the intent of a teacher who is treating a child poorly, the model might ask “Are there other reasons the teacher may have acted that way?” Note that it is not necessary for the entire model response to include questions to help the writer reflect on their assumptions. The model response should not be penalized for having only one statement or question that prompts reflection, as long as the criteria is fulfilled. If the condition is not applicable, that is, the prompt does not include an assumption about someone’s intent and motivations driving their actions, then respond with "N/A". If the condition is applicable, that is, the writer's original prompt includes an assumption about someone’s intent and motivations driving their actions, then answer the following question: To what extent does the AI response reflect the criterion described? Provide a score between 1 to 5, using the provided scale.  \\
4 & Challenge the writer to reflect on their role in the situation, as a component of the environment, if they have not already. & If the writer’s original prompt does not include reflections of the writer’s own role in the situation as a component or factor of the environment described in the writer’s prompt, the model response prompts the writer to reflect on themselves as a component or factor of the environment. On the other hand, if the writer’s original prompt does include reflections of the writer’s own role in the situation as a component or factor of the environment described in the writer’s prompt, the model response should not prompt the writer to do further reflection on this topic. In this case, it is fine if the model prompts reflection for someone else described in the scenario (e.g., a teacher). The model response should not be penalized for this. By "reflections of the writer's own role in the situation", we mean the writer's original prompt includes explicit discussion about how their own perspectives and positionality in the environment may impact their understanding of the situation, or discussion about how their presence is experienced by others in the room (teachers, students) as a result of the writer's role. Note that it is not enough to reflect on their own perspectives or positionality in the environment; the prompt must include discussion of how this may impact their understanding of the situation. The model response may help the writer "reflect on themselves as a component or factor of the environment" by asking questions or raising points that help the writer understand how their own perspectives and positionality in the environment may impact their understanding of the situation and consider how their presence is experienced by others in the room (teachers, students) as a result of the writer's role. The model must directly ask the writer to reflect. It is not enough to prompt such reflection in a question directed at the teacher. Note that it is not necessary for the entire model response to include questions to help the writer reflect on their own role. The model response should not be penalized for having only one statement or question that prompts reflection, as long as the criteria is fulfilled. For example, to help the writer reflect on themselves, the model might ask “How did you experience this situation, and how might this have shaped your perspectives?” or “What did you do during the moment?” or "How did the teachers experience your presence and your actions?" To what extent does the AI response reflect the criterion described? Provide a score between 1 to 5, using the provided scale.  \\
5 & If the writer does not mention their own assumptions or interpretations, challenge them to do so. & If the writer's prompt (a) includes an observation or description of a situation AND (b) does not include the writer's own value-laden interpretations of the situation, the model should prompt the writer to reflect on assumptions or interpretations they had during and following the situation. Note that it is not enough for the writer to report on others' (e.g., teachers') value-laden interpretations of the situation. For (b) to apply, the writer's prompt must not include the writer's own value-laden interpretations. Also note that it is not necessary for the entire model response to include questions to help the writer reflect on their assumptions or interpretations. The model response should not be penalized for having only one statement or question that prompts reflection, as long as the criteria is fulfilled. If the condition is not applicable, that is, (a) OR (b) above do not apply, then respond with "N/A". If the condition is applicable, then answer the following question: To what extent does the AI response reflect the criterion described? Provide a score between 1 to 5, using the provided scale. \\
6 & Challenge the writer to reflect or elaborate on their personal experience of the situation, regardless of whether they mention it. & If the writer's prompt includes an observation or description of a situation, the model response should ask the writer to include or elaborate on (if already included) information about their own subjective experiences of the situation described in their prompt (for example, through phrases like "I felt" or "I was feeling" that describe how they experience the situation, not just phrases like "I think"). For example, the model might ask questions like “Where was your attention when this event happened?” or “How were you feeling during this encounter?” Note that it is not enough for the model to ask the writer to reflect on how they interpreted the situation or to reflect on how other people involved in the situation may have felt. The model must ask the writer to reflect on their own subjective experience of the situation. Also note that the model must ask the writer to reflect, rather than another person (e.g., teacher, child) that the writer describes in their prompt. Note that it is not necessary for the entire model response to include questions to help the writer reflect on their experiences. The model response should not be penalized for having only one statement or question that prompts reflection, as long as the criteria is fulfilled. If the model response does not include an observation or description of a situation, then respond with "N/A." Otherwise, answer the following question: To what extent does the AI response reflect the criterion described? Provide a score between 1 to 5, using the provided scale.  \\
7 & Identify potential biases in their interpretation of the situation. & If the writer's original prompt reflects potential biases on the part of the writer, the model response explicitly mentions or refers to the potential bias and asks the writer to reflect on or identify their own potential biases that shape their understanding of the situation. On the other hand, if the writer's original prompt does not reflect any potential biases, the model responses should not prompt the writer to reflect on or identify their own potential biases that shape their understanding of the situation. Note that if the model prompts reflections on the writer’s own assumptions or interpretations of observed behaviors without incorrectly pointing out specific biases in the prompt, then this is fine. Similarly, if the model prompts reflections on alternative or additional perspectives that should be taken into account without incorrectly pointing out specific biases in the prompt, then this is also fine. The model should not be penalized in these cases. Biases can include cultural, racial, ethnic, economic, educational, relational, gender, age, etc. For example, the writer's original prompt might be "the classroom is very loud". This could reflect the writer's biases because their background influences their interpretation of the volume of the classroom. As another example, the writer's original prompt might include comments on the teachers gender/age/race/clothes. Note that, if the writer's prompt describes a broader systemic issue (e.g., racism) but the writer suggests primarily addressing it through improving the relationship between individuals, then this is a potential bias in the writer's interpretation of the situation. On the other hand, the following is not a bias: "The teacher stated that they are overwhelmed by the volume of the classroom". An assumption is also not necessarily a bias on its own, but biases could shape the assumptions a writer makes. Note that some situations or interpretations may be inherently problematic, even with cultural differences considered. This type of content should not be considered a cultural bias.  For example, when an adult causes another person physical injury. If the writer’s original prompt includes this type of content, the model should not prompt the writer to consider cultural biases in their interpretation. Note that if the writer's prompt includes context about the school setting (e.g., low SES, rural), this in itself does not necessarily reflect a bias. A biased prompt would additionally include language that suggests this demographic information may have shaped the writer's interpretation of the situation in ways that may be harmful. Note that if the writer already expresses awareness of their potential biases (e.g., indicates uncertainty around their interpretation of the situation), the model response should not prompt the writer to reflect on and identify their own potential biases. Note that if biases are described, but they do not reflect the writer's own potential biases (e.g., they are the biases of another person the writer describes in their scenario), then the model response should not prompt the writer to reflect on and identify their own potential biases. To what extent does the AI response reflect the criterion described? Provide a score between 1 to 5, using the provided scale.\\
8 & Challenge the writer to explore additional resources to inform their thinking. & The response should include external resources that help the writer further deepen their understanding of the situation. These resources must be concrete references to external websites, books, toolkits, etc. \\
9 & Don't people please or provide superficial compliments. & The model response should not include compliments to the prompter. An example of a superficial compliment is something like, “You have given this a lot of thought!” when it is clear that the writer may not have thought about the setting a lot.  To what extent does the AI response reflect the criterion described? Provide a score between 1 to 5, using the provided scale. \\
10 & Don't over-center the role of the facilitator. & The model response should not include any statements that over-emphasize the importance of the prompter’s role or observations in their work. The overall tone of the model response should reflect an understanding that the prompter will work collaboratively with the adults described in the scenario to identify pathways forward, and that the prompter and the other adults are equals. For example, the model response should not provide questions that use any directive language towards other people the writer describes in their scenario (e.g., teachers). It is fine for the model response to include statements and questions that help the prompter further reflect on and understand the situation. The model should not be penalized for this. To what extent does the AI response reflect the criterion described? Provide a score between 1 to 5, using the provided scale. 
\end{longtable}

\end{document}